\documentclass[aps,prd,twocolumn,showpacs,showkeys,preprintnumbers,amsmath,amssymb]{revtex4-1}
\usepackage[dvipdfm]{graphicx}
\usepackage[dvipdfm]{rotating}

\begin{document}
\title{On Special Re-quantization of a Black Hole}
\author{S.C. Ulhoa}
\email{sc.ulhoa@gmail.com} \affiliation{Faculdade Gama,
\\Universidade de Bras\'{i}lia, 72405-610, Gama, DF, Brazil.}

\date{\today}

\begin{abstract}
  Quantized expressions for the gravitational energy and momentum are
  derived from a linearized theory of teleparallel gravity. The
  derivation relies on a second-quantization procedure that constructs
  annihilation and creation operators for the graviton. The resulting
  gravitational field is a collection of gravitons, each of which has
  precise energy and momentum. On the basis of the weak-field
  approximation of Schwarzschild's solution, a new form for the
  quantization of the mass of a black hole is derived.
\end{abstract}

\maketitle

\section{Introduction}

One of the most challenging problems in theoretical physics is the
development of a quantum theory of gravitation. General Relativity
successfully describes classical gravitational fields; yet, it is
not a completely satisfactory theory, since it is unable to deal with
certain issues, with dark energy, dark matter, and the nature of
the singularity inside a black hole, for instance~\cite{QG}.

By the Correspondence Principle, a quantum theory of gravity should
reproduce General Relativity in the same way that Quantum
Electrodynamics reproduces Maxwell's equations. It is believed that
such a theory will play an important role at Planck's scale and in a
grand unification scheme, given the extraordinary advance achieved
by unifying three of the four fundamental forces.

An early attempt to quantize gravity is the path-integral method
applied to the Einstein-Hilbert Lagrangian density. In this approach,
only in specific configurations is it possible to make statements
about features of the system; thus, in the case of a black hole the
entropy is found to be proportional to its event-horizon area, which
can be measured by a procedure avoiding space-time
singularities~\cite{PhysRevD.15.2752}. The interaction between
particles inside and outside the event horizon will cause the
evaporation of the black hole; this phenomenon, known as Hawking's
radiation~\cite{springerlink:10.1007/BF02345020}, has motivated recent
experimental investigations~\cite{PhysRevLett.105.203901}.  Whether
such an interaction offers a microscopic explanation for the
black-hole entropy is the subject of debate that has fueled intense
research.

The hamiltonian formulation of General Relativity is constrained,
hence very difficult to deal with. Dirac developed a method to
quantize such systems, which gave rise to a family of theories known
as canonical quantum gravity~\cite{Dirac,CQG}. Among them, Regge
calculus~\cite{springerlink:10.1007/BF02733251,Immirzi:1996dr} and
Loop Quantum Gravity~\cite{lqg,PhysRevD.52.5743} stand out, an
alternative approach being addressed by String Theory.

Every attempt to construct a quantum theory of gravity has to face its
non-renormalizable feature~\cite{springerlink:10.1007/BF00766595} and
deal with the problem of time~\cite{Isham:1992ms}.  Dirac's
quantization method applied to teleparallel gravity could provide an
interesting approach to that goal, since the energy constraint of its
Hamiltonian formulation has shown excellent results when applied to
different systems over the years~\cite{maluf,Ulhoa}.

Loop Quantum Gravity (LQG) is the most prominent theory of quantum
gravity. It tries to quantize gravity in a nonpertubative way,
independently of the chosen spatial background
metric~\cite{lqg,Smolin:2004sx,Ashtekar:2004eh}. The concept of spin
foams being essential, the volume and area operators give rise to
``grains of space-time''. Loop Quantum Cosmology stems from LQG; here,
however, serious conceptual problems have to be addressed, such as the
interpretation of the wave function of the Universe and the avoidance
of the Big Bang singularity~\cite{Barvinsky1992201,Ashtekar:2008zu}.

In grativation, the definition of the energy-momentum vector is very
controversial.  Most attempts lead to coordinate-dependent expressions
or limited ones such as the ADM energy, valid only on the asymptotic
space-time~\cite{ADM}. In the context of teleparallel gravity, by
contrast, a well defined vector arises naturally from the field
equations~\cite{maluf}. This immediately suggests that
second-quantization methods be applied to the gravitational
energy-momentum tensor in a linearized approximation, an approach
adopted in this paper. The gravitational-field properties are then
analogous to those of the electromagnetic field. In particular, the
gravitational energy-momentum tensor being quadratic in the tetrad
field, annihilation and creation operators are easily identified. This
approach contrasts with attempts to quantize matter fields in a
semi-classical approach rooted in Einstein's equations and does not
depend on Dirac's algorithm to identify quantum features of the
gravitational field.

In the following, the greek letters $\mu, \nu, ...$ represent space-time
indices, while the latin letters $a, b, ...$, denote SO(3,1) indices.
In both cases, the indices run from 0 to 3. Time and space indices are
indicated according to the convention $\mu=0,i,\;\;a=(0),(i)$. The
determinant of the tetrad field is denoted $e=\det({e^a}_\mu)$, and
units are such that $G=c=h=1$.

\section{Teleparallel Gravity}

This section presents a brief review of teleparallel gravity. In this theory
the tetrad field, not the metric tensor, is the dynamical field
variable. The tetrad field is related to the metric tensor by the expressions
\begin{eqnarray}
g^{\mu\nu}&=&e^{a\mu}{e_{a}}^{\nu}; \nonumber\\
\eta^{ab}&=&e^{a\mu}{e^{b}}_{\mu}. \label{1}
\end{eqnarray}
where $\eta^{ab}=diag(-+++)$ is the metric tensor of Minkowski
space-time.

Consider now a manifold endowed with the Cartan connection~\cite{Cartan}
$\Gamma_{\mu\lambda\nu}={e^{a}}_{\mu}\partial_{\lambda}e_{a\nu}$,
which can equally well be written in the form
\begin{equation}
\Gamma_{\mu \lambda\nu}= {}^0\Gamma_{\mu \lambda\nu}+ K_{\mu
\lambda\nu}, \label{2}
\end{equation}
where ${}^0\Gamma_{\mu \lambda\nu}$ are the Christoffel symbols, and
$K_{\mu \lambda\nu}$, which is given by
\begin{eqnarray}
K_{\mu\lambda\nu}&=&\frac{1}{2}(T_{\lambda\mu\nu}+T_{\nu\lambda\mu}+T_{\mu\lambda\nu}),\label{3}
\end{eqnarray}
is the contortion tensor defined in terms of the torsion tensor
constructed from the Cartan connection. Thus the torsion tensor is
$T_{\mu \lambda\nu}=e_{a\mu}{T^{a}}_{\lambda\nu}$ with
\begin{equation}
T^{a}\,_{\lambda\nu}=\partial_{\lambda} {e^{a}}_{\nu}-\partial_{\nu}
{e^{a}}_{\lambda}. \label{4}
\end{equation}

The curvature tensor obtained from $\Gamma_{\mu \lambda\nu}$ is
identically zero. Equation~(\ref{2}) thus leads to
\begin{equation}
eR(e)\equiv -e({1\over 4}T^{abc}T_{abc}+{1\over
2}T^{abc}T_{bac}-T^aT_a)+2\partial_\mu(eT^\mu).\label{5}
\end{equation}

We then drop the divergence on the right-hand side to define the
teleparallel Lagrangian density
\begin{eqnarray}
\mathfrak{L}(e_{a\mu})&=& -\kappa\,e\,({1\over 4}T^{abc}T_{abc}+
{1\over 2} T^{abc}T_{bac} -T^aT_a) -\mathfrak{L}_M\nonumber \\
&\equiv&-\kappa\,e \Sigma^{abc}T_{abc} -\mathfrak{L}_M\;, \label{6}
\end{eqnarray}
where $\kappa=1/(16 \pi)$, $\mathfrak{L}_M$ stands for Lagrangian
density of matter fields, and $\Sigma^{abc}$ is given by
\begin{equation}
\Sigma^{abc}={1\over 4} (T^{abc}+T^{bac}-T^{cab}) +{1\over 2}(
\eta^{ac}T^b-\eta^{ab}T^c)\;, \label{7}
\end{equation}
with $T^a={{T^b}_b}^a$. It is important to notice that the
Einstein-Hilbert Lagrangian density is equivalent to its teleparallel
version (\ref{6}). Thus, concerning observational data, both theories
share the same results.

We then compute the variational derivative of the Lagrangian density with
respect to $e^{a \mu}$ to obtain the following field equations:
\begin{equation}
e_{a\lambda}e_{b\mu}\partial_\nu(e\Sigma^{b\lambda \nu})-
e({\Sigma^{b \nu}}_aT_{b\nu \mu}- {1\over
4}e_{a\mu}T_{bcd}\Sigma^{bcd}) \;= {1\over {4\kappa}}eT_{a\mu},
\label{8}
\end{equation}
where $T_{a\mu}$ is the energy-momentum tensor of matter
fields. Explicit calculation shows that Eq.~(\ref{8}) is equivalent to
Einstein's equations.

The field equations can be rewritten in the form
\begin{equation}
\partial_\nu(e\Sigma^{a\lambda\nu})={1\over {4\kappa}}
e\, {e^a}_\mu( t^{\lambda \mu} + T^{\lambda \mu}), \label{10}
\end{equation}
where $t^{\lambda\mu}$ is defined by the equality
\begin{equation}
t^{\lambda \mu}=\kappa(4\Sigma^{bc\lambda}{T_{bc}}^\mu- g^{\lambda
\mu}\Sigma^{bcd}T_{bcd}). \label{11}
\end{equation}
Since $\Sigma^{a\lambda\nu}$ is skew-symmetric in the last two
indices, it follows that
\begin{equation}
\partial_\lambda\partial_\nu(e\Sigma^{a\lambda\nu})\equiv0.\label{12}
\end{equation}

We thus see that
\begin{equation}
\partial_\lambda(et^{a\lambda}+eT^{a\lambda})=0,\label{13}
\end{equation}
which yields the continuity equation
\begin{equation*}
{d\over {dt}} \int_V \mathrm{d}^3x\,e\,{e^a}_\mu (t^{0\mu} +T^{0\mu})
=-\oint_S dS_j\, \left[e\,{e^a}_\mu (t^{j\mu} +T^{j\mu})\right].
\end{equation*}

Accordingly, $t^{\lambda \mu}$ is interpreted as the energy-momentum tensor of
the gravitational field~\cite{maluf2}, and the total energy-momentum in a
three-dimensional volume $V$ of space is
\begin{equation}
P^a = \int_V \mathrm{d}^3x \,e\,{e^a}_\mu(t^{0\mu}+ T^{0\mu}). \label{14}
\end{equation}
The right-hand side remains invariant under coordinate transformations,
transforms like a vector under Lorentz transformations and hence displays
the expected features of a true energy-momentum vector.

\section{Quantized Energy and Re-quantization of Matter}
We now turn our attention to a linearized theory of teleparallelism. In this
approximation the tetrad field can be written in the form
\begin{equation}
{e^a}_{\mu}\approx\delta^{a}_{\mu}+{\psi^{a}}_{\mu},\label{15}
\end{equation}
and the metric tensor has a similar expansion:
\begin{equation*}
  g_{\mu\nu}=\eta_{\mu\nu}+h_{\mu\nu}.
\end{equation*}
At this point, the energy-momentum of matter fields is zero, and the
energy-momentum vector $P^a$ depends on $t^{a\mu}$ only. Simple
algebraic manipulations convert $t^{a\mu}$ to the form
\begin{small}
\begin{eqnarray}
t^{0a}&\approx&-\kappa\frac{1}{4}\delta^a_{\mu}(\partial^{\nu}\psi^{c0}
-\partial^{0}\psi^{c\nu})\partial^{\mu}\psi_{c\nu}
-\kappa\frac{1}{2}\delta^{a}_{\mu}\eta^{0\mu}\times\nonumber\\
&&\times\Big((\partial_{\nu}{\psi^c}_{\gamma})(\partial^{\gamma}{\psi_c}^{\nu})
-(\partial^{\lambda}{\psi^c}_{\lambda})(\partial_{\mu}{\psi_c}^{\mu})\Big).\label{16}
\end{eqnarray}
\end{small}

Consider now the Fourier expansion of $\psi_{a\mu}$ in a three-dimensional
normalized volume,
\begin{equation}
\psi_{a\mu}=\frac{1}{\sqrt{2\omega}}\sum_{\mathbf{k}}
\Big(A_{a\mu}(\mathbf{k},t)e^{\imath\mathbf{k}\cdot\mathbf{x}}
+A^{\dag}_{a\mu}(\mathbf{k},t)e^{-\imath\mathbf{k}\cdot\mathbf{x}}\Big),\label{17}
\end{equation}
where the time dependence of the coefficients $A_{a\mu}(\mathbf{k},t)$
is of the form $e^{-\imath\omega t}$. For the purposes of second
quantization, it is sufficient to consider terms containing
$A^{a\mu}(\mathbf{k},t)A^{\dag}_{a\mu}(\mathbf{k},t)$ and analogous
combinations, no generality being lost, since
quadratic terms with $A^{a\mu}(\mathbf{k},t)$ and
$A^{\dag}_{a\mu}(\mathbf{k},t)$ do not contribute to the hamiltonian
operator in the context of annihilation and creation
operators. Notice, moreover, that the time average of these quadratic
terms over a period vanishes. Thus, in the transversal
gauge $k^{\mu}\psi_{a\mu}=0$, where $k^{\mu}=(\omega,\mathbf{k})$ with
$k^{\mu}k_{\mu}=0$, the time average of the energy in Eq.~(\ref{14}) reads
\begin{equation}
P^{(0)}\approx\kappa\sum_{\mathbf{k}}\frac{1}{2}\omega
\Big(A^{a\mu}(\mathbf{k})A^{\dag}_{a\mu}(\mathbf{k})+A^{\dag a\mu}(\mathbf{k})
A_{a\mu}(\mathbf{k})\Big).\label{18}
\end{equation}
The coefficients in (\ref{17}) assumed to be operators, the constraint
$\lbrack A^{\dagger a\mu}({\bf k}), A_{b\nu}({\bf k}^\prime)\rbrack=
\delta_{{\bf k}{\bf k}^\prime} \delta^a_b \delta^\mu_\nu$ leads to the
hamiltonian operator

\begin{equation}
\hat{H}\approx\kappa\sum_{\mathbf{k}}\frac{1}{4}\omega\Big(\hat{A}^{\dag
a\mu}(\mathbf{k})\hat{A}_{a\mu}(\mathbf{k})+\frac{1}{2}\Big),\label{19}
\end{equation}
where the $\hat{A}^{\dag a\mu}(\mathbf{k})$ are creation and the
$\hat{A}_{a\mu}(\mathbf{k})$ are annihilation operators. A similar
analysis starting from Eq.~(\ref{14}) leads to the following
expression for the gravitational momentum in terms of creation and
annihilation operators:
\begin{equation}
\hat{\mathbf{P}}\approx\kappa\sum_{\mathbf{k}}\frac{1}{4}\mathbf{k}\Big(\hat{A}^{\dag
a\mu}(\mathbf{k})\hat{A}_{a\mu}(\mathbf{k})+\frac{1}{2}\Big).\label{20}
\end{equation}

Equations~(\ref{19})~and (\ref{20}) are remarkably similar to the QED
expressions for the electromagnetic energy and momentum,
respectively. If a set of orthonormal wave functions obeying
$\hat{H}\Psi=\mathcal{E}\Psi$ is postulated, then the eigenvalue of
$\hat{H}$ is
\begin{equation}\label{eq:1}
\mathcal{E}\approx\kappa\sum_{\mathbf{k}}\frac{1}{4}\omega
N_{\mathbf{k}},
\end{equation}
where $N_{\mathbf{k}}$ is the number of gravitons, and the constant
second term within parentheses on the right-hand side of
Eq.~(\ref{19}) was dropped to avoid a trivial divergence.  In this
context, the gravitational field can be regarded as a collection of
gravitons playing the role of photons in an electromagnetic field.

In order to make a prediction, let us consider the space-time
around  a black hole, which is described by the line element
\begin{small}
\begin{equation}
ds^2=-(1-2M/r)dt^2+(1-2M/r)^{-1}dr^2+r^2d\theta^2+r^2\sin^2\theta
d\phi^2,\label{21}
\end{equation}
\end{small}
found in Schwarzschild's solution~\cite{Dinverno}. Under the weak
field condition $M/r<<1$, Eq.~(\ref{21}) fits our linearized theory.

For stationary observers, Eq.~(\ref{14}) is reduced to the form
\begin{equation*}
  \mathcal{E}=\frac{1}{8\pi}\int_S \mathrm{d}\theta\, \mathrm{d}\phi (eT^1),
\end{equation*}
where
\begin{equation*}
  eT^1=2r\sin\theta\left(1-\sqrt{\left(1-\frac{2M}{r}\right)}\right).
\end{equation*}

If we then let $S \rightarrow \infty$, to cover the whole space-time,
we find that the energy of the Schwarzschild space-time is
$\mathcal{E}=M$, a result that is well known in the context of
asymptotic expressions such as the ADM energy (see for instance
\cite{Maluf:1995re}) and holds equally well in the weak-field
approximation. We can therefore extract the mass of the black hole
from Eq.~(\ref{eq:1}), which yields the expression
\begin{equation}
M\approx\kappa\sum_{\mathbf{k}}\frac{1}{4}\omega
N_{\mathbf{k}}.\label{22}
\end{equation}

This identification is valid only for inertial frames, since it
depends on our derivation of the energy-momentum vector from the
expansion~(\ref{15}), and the chosen gauge. Notice should be taken
that Eq.~(\ref{22}) was derived in an empty space-time and
relates the number of gravitons to the mass of the black hole. That
result tells us that mass is quantized in a new sense, other than the
concept of particle or atom. For this reason, we refer to it as a
requantization of a black hole.

Since the mass of the black hole on the left-hand side of
Eq.~(\ref{22}) is finite, the right-hand side is expected to be finite
as well. The sum should therefore be truncated at some point, and the
mass of the black hole, given by
\begin{equation*}
M\approx m_0+m_1+\ldots+m_N
\end{equation*}
where each $m$ is a quantum of mass dependent on a specific frequency.

That the gravitons are bosons is shown by the commutation relations
between the annihilation and creation operators and by the eigenvalue
of the hamiltonian operator.  Although the preceding analysis
has led to no conclusion concerning spin, preliminary studies
extending the teleparallel-gravity procedure to the gravitational angular
momentum suggest that the graviton has spin 2.

\section{Conclusion}

The results in this paper can be summarized as follows. Teleparallel
gravity yields an energy-momentum vector that is independent of
coordinate transformations. In a linearized theory, $\psi_{a\mu}$ can
be expanded in a Fourier series, and this leads to second-quantized
expressions for the energy and momentum. For inertial frames, the energy of
Schwarzschild's space-time is given by $M$, a result that is valid
even in the weak-field approximation and identifies a new form of
quantization, which is associated with the number of gravitons.

It should be noticed that the requantization encompasses only the
gravitational energy-momentum; no matter fields are involved.  The
resulting new quantization of the font of gravitation contrasts with
what Loop Quantum Gravity asserts with its discrete space-time.

The definition of the ADM energy shows that the mass of a black hole
has to be associated with a particle of integer spin\textemdash the
graviton.  Matter, however, is constituted of fermions. The resulting
apparent contradiction constitutes a serious obstacle obstructing the
development of quantum theories of gravity.

The next challenge, thus, is to develop a general theory for the
requantization of mass, one that identifies general creation and
annihilation operators. A general equation for quantum gravity in the
teleparallel case could be analogous to the Wheeler-DeWitt
equation~\cite{PhysRev.160.1113,PhysRev.162.1195,PhysRev.162.1239},
and hence bring to light the quantum nature of matter. The pursuit of a
quantum theory of gravity should not treat matter and space-time
on different footings. The inextricable connection between the two
entities should emerge from the (re)quantization of matter itself.

\bibliography{ms1956}

\end{document}